
%
%
%
%
\def\etal{\it et al. \rm}
\def\kms{km s$^{-1}$}

\def\hmpc{{\rm h}$^{-1}$ {\rm Mpc}}

\def\gsim{ \lower .75ex \hbox{$\sim$} \llap{\raise .27ex \hbox{$>$}} }
\def\lsim{ \lower .75ex \hbox{$\sim$} \llap{\raise .27ex \hbox{$<$}} }
\def\pp{\noindent\parshape 2 0truecm 16.0truecm 1truecm 15truecm}
\def\pp{\noindent\parshape 2 0truecm 15truecm 2truecm 13truecm}

\def\spose#1{\hbox to 0pt{#1\hss}}
\def\simlt{\mathrel{\spose{\lower 3pt\hbox{$\mathchar"218$}}
     \raise 2.0pt\hbox{$\mathchar"13C$}}}
\def\simgt{\mathrel{\spose{\lower 3pt\hbox{$\mathchar"218$}}
     \raise 2.0pt\hbox{$\mathchar"13E$}}}

\hrule height0pt
\magnification=\magstep1
\baselineskip 12pt
\parskip=6pt
\parindent=0pt
\hsize=6.5truein
\vsize=8.6truein


\font\titlefont=cmss17

\centerline{\titlefont THE \ CLUSTERING \ OF \ IRAS \ GALAXIES}

\vskip 0.8truein
\centerline{\bf Ben Moore$^1$, Carlos S. Frenk$^2$, George Efstathiou$^3$,}
\vskip 0.12truein
\centerline{\bf and Will Saunders$^3$}
\vskip 0.3truein

\centerline{\it 1. Department of Astronomy, University of California,
Berkeley, CA94720, USA}
\centerline{\it 2. Department of Physics, University of Durham, DH1 3LE,
England}
\centerline{\it 3. Department of Physics, University of Oxford, 0X1 3RH,
England}

\vskip 0.4truein
\centerline {\bf ABSTRACT}
\vskip 8pt
\parindent=36pt

We investigate the clustering of galaxies in the QDOT redshift survey of
IRAS galaxies. We find that the two-point correlation function is well
approximated by a power-law of slope $-1.11\pm0.09$, with clustering
length $3.87\pm0.32 {\rm h}^{-1}$ Mpc\footnote*{Throughout this paper $h$
denotes the Hubble constant in units of 100 ${\rm km}{\rm s}^{-1}{\rm
Mpc}^{-1}$.} out to pair separations of $\sim 25 {\rm h}^{-1}$ Mpc. On
scales larger than $\sim 40 {\rm h}^{-1}$ Mpc, the correlation function is
consistent with zero. The {\it rms} fluctuation in the count of QDOT
galaxies above the Poisson level in spheres of radius $8 {\rm h}^{-1}$ Mpc
is $\sigma_8^{IRAS}=0.58\pm0.14$, showing that fluctuations in the
distribution of IRAS galaxies on these scales are smaller than those of
optical galaxies by a factor of about $0.65$, with an uncertainty of $\sim
25 \%$. We find no detectable difference between the correlation functions
measured in  redshift space and in real space, leading to a $2\sigma$
limit of $b_{IRAS}/\Omega^{0.6} >1.05$, where $b_{IRAS}$ is the bias
factor for IRAS galaxies and $\Omega$ is the cosmological density
parameter. The QDOT autocorrelation function calculated in concentric
shells increases significantly with shell radius.  This difference is more
likely due to sampling fluctuations than to an increase of the clustering
strength with galaxy luminosity, but the two effects are  difficult to
disentangle; our data allow at most an increase of $\sim 20 \%$ in
clustering strength for each decade in luminosity. For separations greater
than  $\sim 3 {\rm h}^{-1}$ Mpc, the cross-correlation function of Abell
clusters (with Richness $R\ge 1$) and QDOT galaxies is well approximated
by a power-law of slope $-1.81\pm0.10$, with clustering length
$10.10\pm0.45 {\rm h}^{-1}$ Mpc, and no significant signal beyond $\gsim
50 {\rm h}^{-1}$ Mpc. This cross-correlation depends only weakly on
cluster richness.

\vskip 0.4truecm


\vfill\eject

\vskip 20pt
\centerline {\bf \S 1. INTRODUCTION}
\vskip 20pt
\parindent=36pt

The distribution of matter in the Universe on large scales provides one of
the strongest constraints on models of galaxy formation. Several surveys
have been used to measure galaxy clustering on scales between  $ 10 {\rm
h}^{-1}$ Mpc and $50 {\rm h}^{-1}$ Mpc  (Maddox \etal 1990 (APM); Efstathiou
\etal 1990, Saunders \etal 1991 (QDOT); Fisher \etal 1993 (1.2Jy); Vogeley
\etal 1992 (CfA2); Loveday \etal 1992a (Stromlo/APM)). In addition,
estimates of the mass and abundance of galaxy clusters (White, Efstathiou
and Frenk 1993) and measurements of large scale velocity flows (Vittorio
\etal  1986, Lynden-Bell \etal 1988, Strauss \etal 1993) have been used to
constrain  the mass  distribution on scales of $\sim 10$ and up to $\sim 100
{\rm h}^{-1}$ Mpc respectively. On scales comparable to the present Hubble
radius, temperature fluctuations in the microwave background can be used as
probes of primordial fluctuations in the gravitational potential (Smoot
\etal 1992).

In this paper, we analyse the clustering properties of the
QDOT ``one-in-six" redshift survey of IRAS galaxies (Lawrence \etal 1993).
There are $2163$ galaxies in this survey which were chosen by randomly
sampling at a rate of one-in-six from galaxies selected from the
IRAS point source
catalogue with $60\mu$ flux greater than 0.6Jy at galactic latitudes
$|b|>10^\circ$. Regions of sky heavily contaminated by sources in our own
galaxy and the IRAS coverage gaps are delineated by a mask which excludes
about 10\% of the sky at $|b|>10^\circ$. The parent galaxy catalogue is
described in detail by Rowan-Robinson \etal (1991). The QDOT survey covers
$9.31$ steradians (74\% of the sky) and samples the density field in the
Universe to a redshift $z\approx0.05$, well beyond the Local Supercluster.
The survey is particularly well suited to investigations of large scale
structure because it uniformly samples a large-volume of the universe.

A variety of techniques have been used to characterise large-scale
clustering  in the QDOT survey.  Efstathiou \etal (1990) and Saunders
\etal (1991) estimated the variance of QDOT galaxy counts-in-cells
and found that the clustering on scales $\simgt 20$\hmpc\ exceeds that
expected in simple versions of the cold dark matter (CDM) model. Moore \etal
(1992) studied the topology of the galaxy distribution and showed that the
power spectrum is consistent with that of a Gaussian random field with the
slope inferred from the counts-in-cells analyses.  These studies
complement each other and give consistent results, even though
the estimation techniques and weighting schemes are different.

The analysis of IRAS galaxy correlations presented here is
directed at three distinct questions:

\noindent
[A] Although the above analyses of the QDOT sample have revealed a
discrepancy with the standard CDM model, they assume that the galaxy
luminosity function is universal and independent of position and local
environment. Babul and White (1991), Silk (1992) and Bower \etal (1993)
have considered alternative non-linear biasing schemes which could give
rise to enhanced large scale power in the galaxy distribution. Such
effects, if present, complicate the interpretation of clustering data and
could remove the discrepancy with the standard cold dark
matter model. Spatial modulations in the efficiency of galaxy formation
could be tested by measuring the two-point correlation function as a
function of luminosity or morphological type. In this paper we investigate
how the correlation function of IRAS galaxies depends on luminosity.

\noindent
[B] Saunders \etal (1992) cross-correlated the QDOT redshift catalogue
with its parent IRAS catalogue, which contains the positions of over 12000
galaxies, to compute a cross-correlation function as a function of metric
separation. This function is related to the spatial correlation function,
$\xi(r)$, by a Limber-like equation which can be inverted numerically.
Comparing this determination of $\xi(r)$  with the redshift-space
correlation function measured in this paper yields information about the
clustering distortion caused by the peculiar velocities of galaxies and
hence about the relative clustering strength of galaxies and mass on large
scales (Kaiser 1987).

\noindent
[C] Several authors have pointed out that IRAS galaxies appear more weakly
clustered than optically selected galaxies on small scales (e.g. Davis
\etal 1989; Saunders \etal 1992; Strauss \etal 1992). In addition, the
cores of rich clusters appear to contain very few IRAS galaxies. Since these
are predominantly late type disk systems, these observations may simply
reflect the morphology-density relation. In an attempt to understand their
environments, we study how IRAS galaxies are distributed with respect to
rich Abell clusters by cross-correlating the positions of the QDOT galaxies
with a sample of rich galaxy clusters. The nature of the IRAS luminosity
function creates a tail of high redshift QDOT galaxies and allows us to
directly measure this relation in redshift space. We compare these results
with those of Lilje and Efstathiou (1988) who measured the projected
cross-correlation function between optically selected galaxies and Abell
clusters.

In the next section we calculate the two-point correlation function of the
QDOT redshift survey. We compare different weighting schemes and methods to
determine the errors in $\xi(s)$ using artificial IRAS  catalogues
constructed from numerical simulations of the CDM model.  From Monte-Carlo
simulations and bootstrap resampling, we show how uncertainties in estimates
of the correlation function can be reliably obtained from the data.
In Section 3 we compare the redshift-space and real-space correlation functions
from numerical simulations and the QDOT survey in an attempt to quantify the
distortions produced by large-scale streaming motions.
In Section 4 we subdivide the data by distance and luminosity to investigate
the
large scale clustering properties of these galaxies and to test for luminosity
dependent clustering. In Section 5 we cross-correlate the QDOT data with
samples of Abell clusters of different richness.
Our results are summarised in Section 6 where we also
compare with previous analyses of these data.

\vskip 0.2 truein
\noindent
{\bf \S 2. THE TWO-POINT CORRELATION FUNCTION}

\vskip 0.2 truein

The two-point correlation function of a redshift survey, $\xi(s)$, is
defined in terms of the joint probability of finding galaxies separated by
distance $s$ in redshift space and lying in volume elements $\delta V_1$ and
$\delta V_2$:
$$
\eqalignno{
 \delta P_{12} & = \overline n ^2 (1 + \xi(s)) \delta V_1 \delta V_2 \ ,
&(1)\cr
}
$$
where $\overline n$ is the mean density. To calculate $\xi(s)$ for the QDOT
survey we use two different estimators, one introduced by Davis and Peebles
(1982) and the other by Hamilton (1993). The first of these is:
$$
\eqalignno{
{1 + \xi(s)} &= {DD(s) \over DR(s)} {\biggl( {N_R\over N_D} \biggr)} \ ,
&(2a)\cr
}
$$
where $DD(s)$ is the weighted pair count of galaxies in an interval of
separation centred on $s$ and $DR(s)$ is an analogous weighted count of
pairs consisting of a galaxy and a point chosen from a large random
catalogue with the same sky coverage and mean
redshift distribution as the QDOT survey. $N_R$ and $N_D$ are weighted
sums over the random and data points given by equation (5) below.
Hamilton (1993) has proposed a slightly different estimator which
is less sensitive to uncertainties in the mean density:
$$
\eqalignno{
{1 + \xi(s)} &= {DD(s)RR(s) \over DR(s)DR(s)} \ ,
&(2b)\cr
}
$$
where $RR(s)$ is the weighted count of pairs of points in
the random catalogue.

For both these estimators the number of data-data,
data-random and random-random pairs are calculated as
$$
\eqalignno{
DD, DR \ or \ RR &= \sum_i \sum_j {1\over (1+4\pi J_3(s_{ij}) \phi(s_i))}
{1\over (1+4\pi J_3(s_{ij}) \phi(s_j))} \ ,
&(3)\cr
}
$$
where the sum over $i$ extends over the entire sample and the sum over
$j$ only over those sample members within the specified distance bin
from the $i^{\rm th}$ member; $\phi(s_i)$ is the mean galaxy density at
distance
$s_i$; $s_{ij}$ is the separation, $\vert \underline s_i- \underline s_j
\vert$, between  the $i^{\rm th}$ and $j^{\rm th}$ sample members and
$$
\eqalignno{
J_3(s_{ij}) &= \int_0^{s_{ij}} s^2\xi(s) ds \ .
&(4)\cr
}
$$
The weights in equation (3) are chosen to give an
approximately minimum variance estimate
of $\xi(s)$ on large scales ({\it c.f.} Efstathiou 1988, Loveday \etal
1992a, Saunders \etal 1992). This weighting scheme has the desirable property
of switching from  volume weighting at small distances where
clustering dominates the error in $\xi$, to uniform weighting per galaxy at
large separations, as is appropriate for a Poisson point process.

The number densities are calculated as
$$
\eqalignno{
N_D(s) &= 1/{V_w}\sum_{data,i}w_i(s) \ , \ \ \
N_R(s) = 1/{V_w}\sum_{random,i}w_i(s) \ ,
&(5)\cr
}
$$
where
$$
\eqalignno{
w_i &= {1\over 1+4\pi J_3^{max} \phi(s_i)} \ ,
&(6)\cr
}
$$
and $J_3^{max}$ is the maximum value of the integral
$J_3$ defined in equation~(4) and is derived
from the models of $\xi(s)$ described below
(see for example, Loveday \etal 1992b).
$V_w$ is the weighted volume:
$$
\eqalignno{
V_w &= 4\pi\int s^2 \phi(s) w(s) ds
 \ .
&(7)\cr
}
$$

For this analysis we require prior knowledge of the galaxy selection
function, $\phi(s)$, and a model for $\xi(s)$ itself. We calculate the
selection function by integrating the luminosity function of IRAS galaxies
(solution (19) of Saunders \etal 1990). Since the fractional error in
$1+\xi(s)$ is proportional to $\Delta\phi/\phi$, uncertainties in the mean
density at distances $\gsim 100 {\rm h}^{-1}$ Mpc can lead to errors in
$\xi(s)$ of similar size to the signal that we are trying to measure.
Saunders \etal (1991) determined the mean density of the QDOT survey to
within $\pm 5\%$, an uncertainty which arises mainly from small number
statistics and systematic effects due to density fluctuations of order the
survey size. Furthermore, Saunders \etal (1990) found evidence for evolution
in the number density of IRAS galaxies even at relatively low redshifts,
$n\propto (1+z)^{6.7 \pm 2.3}$. The uncertainty in this estimate could
systematically bias our estimates of the correlation function, so in what
follows we explicitly include the effects of evolution. We find that for
the most part these effects are negligible.

To compute $J_3$ for the weighting scheme of equation (3) we use the linear
theory correlation function for models similar to the standard scale-invariant
CDM model. The shapes of the power-spectra in these models are characterised by
a parameter, $\Gamma$, which is simply $\Gamma=\Omega h$ in adiabatic, scale
invariant CDM models with low baryon density. Lower values of $\Gamma$
correspond to models with more power on large scales (see Efstathiou, Bond and
White (1992) for a discussion of constraints on the parameter $\Gamma$). We
consider the values $\Gamma = 0.5\ {\rm and}\ 0.2$. These models are
normalised such that the variance of the density field in top hat spheres of
radius  $8 {\rm h}^{-1}$ Mpc equals 0.7, consistent with the counts-in-cells
results of Efstathiou \etal (1990) and Saunders \etal (1991).

To test different weighting schemes and other statistics, we constructed
artificial QDOT catalogues from the ensemble of 5 CDM N-body simulations
described by Frenk \etal (1990). At the output time chosen, the rms mass
fluctuation in spheres of radius $8 {\rm h}^{-1}$ Mpc, $\sigma^{\rho}_8=
0.5$, consistent with the model independent determination of
$\sigma_8^{\rho}$ by  White, Efstathiou and Frenk  (1993). Using the
techniques introduced by White \etal (1987), based on the ``high-peak"
model of biased galaxy formation, we identified galaxies in the
simulations and constructed catalogues with the same flux limit,
sampling rate, masked region and  mean  selection function of the QDOT
survey. The average (real-space) rms fluctuation in the galaxy count in
these catalogues is  $\sigma_8^{g}=0.76$, corresponding to a biasing
parameter, $b\equiv \sigma_8^{\rho}/\sigma_8^{g}=1.5$. This is only
slightly higher than the value inferred for QDOT  galaxies in an
$\Omega=1$ universe by Kaiser \etal (1991) from an analysis of peculiar
velocity fields. In all subsequent calculations we limit our samples to
distances less than 300\hmpc\ and we use random  catalogues containing
$10^5$ galaxies.

Figure~$1a$ shows mean values of the redshift-space $\xi(s)$ (ie using the
redshift-distance rather than the true distance to each galaxy) for the
artificial catalogues, estimated from equation (2a).
Results with three different weighting schemes are displayed:  equation~(3)
with $J_3$ obtained by integrating the  linear theory prediction for
$\xi(s)$ with $\Gamma=0.5$ and $0.2$;  equation~(3) assuming a constant
$4\pi J_3=3000$; and equal weighting for all galaxies. The dashed line shows
the linear theory prediction for the CDM ($\Gamma=0.5$) correlation
function, corrected for redshift-space distortions as discussed in
Section~3.

Our estimates of $\xi(s)$ do not depend strongly on our assumed weighting
scheme even on large scales. Furthermore, on large scales the estimates from
the simulations agree very well with the linear theory prediction. Figure 1b
shows the mean of $\xi(s)$ estimated from equation~(2a) (with $J_3$
from linear theory), together with 1$\sigma$ errors determined from
the scatter in the 5 different realisations. It is
interesting to see how reliably we can estimate the errors in $\xi(s)$ from
a single sample. To this effect we plot to the left of each point in Figure
1(b) the statistical errors, $\Delta\xi = (1+\xi(s))/\sqrt{N_{D}}$, and to
the right of each point, the $1\sigma$ errors from $100$ bootstrap
realisations of a single artificial catalogue. The different error estimates
give similar results with no measurable bias. Since the ($J_3$-weighted)
statistical errors are the simplest to calculate, we shall use them in the
remainder of this paper.

Figure~2a shows the correlation function of the QDOT survey calculated from
equation~(2b), with the $\Gamma=0.2$ linear theory prediction for
$J_3(s_{ij})$. The open and filled circles show the effect of including and
ignoring evolution in the number density with redshift. Our estimates of the
correlation function are very stable, even though the density of galaxies in
the random catalogue at 100\hmpc\ (approximately the median distance)
increases by $\sim$ 25\% when evolution is included. Figure~2b shows
correlation functions, this time estimated using equation~(2a). (The symbols
are as in Figure~2a.) Both estimators give very similar results.

The data in Figures~2a and 2b can be approximately described by a power-law
on scales $\lsim 25$\hmpc, as expected from the angular correlation function
of IRAS galaxies which is very well fit by a power-law on scales 0.1 to 5
${\rm h}^{-1}$ Mpc. The small curvature apparent in the QDOT data can be
explained by the combined effect of peculiar velocities and observational
errors in the velocity determinations. Both of these enhance the
correlations on scales of a few hundred kilometers per second and suppress
them on smaller scales. Ignoring for the moment the effect of large-scale
galaxy bulk flows, we can model these small-scale effects by
assuming a form for the distribution of relative velocities between galaxy
pairs. If $\xi_r^\prime$ denotes the spatial correlation function in real
space (unaffected by velocity errors and peculiar velocities on small
scales), then the observed correlation function in redshift space will be
the direction-averaged convolution of $\xi_r^\prime(s)$ and the velocity
distribution function, $f(v)$, {\it i.e.}
$$
\eqalignno{
\xi_s(s) &\approx {1\over 2 \pi}\int_{-1}^{1}\int_{-\infty}^{\infty}
\xi_r^\prime(\sqrt{s^2 + v^2 -2sv\mu}) f(v)\;dvd\mu  \ ,
&(8) \cr
}
$$
where we have ignored any dependence of the shape of $f(v)$ on
pair separation and we have ignored distortions caused by
streaming motions (see equation 10 below) which are small if
$\xi \simgt 1$ (see {\it e.g.} Bean \etal 1983).
We assume that $\xi_r^\prime(r)=(r/r_0)^{-\gamma}$ and that $f(v)$ is a
Gaussian
with dispersion $\sigma$. We then find the values of $r_0$ and
$\gamma$ that give the best fit between $\xi_s(s)$ and the data points in
Figure 2a. Adopting $\sigma=250$\kms, a $\chi^2$ fit to the data out to
pair separations of $25{\rm h^{-1} Mpc}$ gives:
$$
\eqalignno{
r_0 &=3.87 \pm 0.32{\rm h^{-1} Mpc}, \ \ \gamma=1.11 \pm 0.09 \ .
& (9)\cr
}
$$
Very similar values are obtained if instead we take $\sigma=400$\kms. Table 1
summarises our
correlation function results. Both $\xi_s(s)$ and $\xi_r^\prime(s)$ are
plotted in Figure~2a.

We have also plotted in Figure~2a the average correlation function in
redshift space for our five artificial catalogues, as well as the  linear
theory prediction of the {\it real-space} correlation function for a model
with $\Gamma=0.2$. (As discussed below, redshift-space distortions are
negligible in the latter case.) On scales $\gsim 10 {\rm h}^{-1}$ Mpc, the
data show the same excess clustering relative to the standard CDM model seen
in previous studies.
The $\Gamma=0.2$ linear theory model
provides a good match to the data on  scales $s\simeq 10 {\rm h}^{-1}$ Mpc.
This model also gives a good match to the shape of the large-scale angular
correlation function of optical galaxies determined from the APM survey
(Maddox \etal 1990). Beyond $s\simeq 40 {\rm h}^{-1}$ Mpc, the QDOT
correlations are consistent with zero.

In Figure~2b, we plot correlation functions for three subsamples of the
QDOT survey, volume-limited at $40$, $80$ and $120$ ${\rm h}^{-1}$ Mpc,
containing $169$ , $295$ and $298$ galaxies respectively. On small scales,
the various
curves are roughly consistent with one another. However, on scales $>10 {\rm
h}^{-1}$ Mpc, $\xi(s)$ estimated from the full survey has a somewhat
larger amplitude than $\xi(s)$ estimated from the first two volume-limited
subsamples. We shall see below that this discrepancy is caused by a small but
systematic increase of clustering amplitude with distance.

\vskip 20pt
\parindent=0pt
{\bf \S 3. COMPARISON OF THE REAL AND REDSHIFT SPACE CORRELATION FUNCTIONS}
\vskip 20pt
\parindent=36pt

We have measured the correlation function of QDOT galaxies in redshift
space. According to linear perturbation theory, streaming motions cause the
redshift-space correlation function, $\xi_s$, to be enhanced over that
measured in real space, $\xi_r$, by a factor
$$
\eqalignno{
 f &= 1 + {2 \over 3} {\Omega_0^{0.6} \over b} + {1 \over 5} {\Omega_0^{1.2}
\over b^2} \ ,
&(10)\cr
}
$$
(Kaiser 1987) where we have assumed that in the linear regime
the correlation function of IRAS
galaxies is related to the mass correlations, $\xi_\rho(r)$, by $\xi_r(r) =
b^2 \xi_\rho (r)$ ({\it i.e.} linear biasing; see Davis \etal 1985).

To test the validity of this formula in the regime probed by our data and to
assess our ability to measure $f$ from the QDOT survey, we first  calculate the
two-point correlation function in real- and redshift-space for our artificial
catalogues.  The filled and open squares in Figure~3 show the average
real-space
and redshift-space correlation functions respectively for our 5 fully-sampled
catalogues. The lower solid line shows a polynomial fit to the correlation
funtion in real space, and the upper curve shows this same polynomial fit
shifted vertically by the expected factor of 1.53 as indicated by equation~(10)
for $b=1.5$. This line gives a surprisingly good match to the redshift-space
correlation function, even on scales where the clustering pattern is mildly
non-linear.

The results for the artificial one-in-six catalogues are less encouraging.
The redshift - space correlation function lies systematically above the
(fully-sampled) real-space function, but the statistical errors are too
large for this effect to be detected with high significance. In practice,
the uncertainties in estimating the real-space correlation function are
likely to exacerbate this problem. An estimate of $f$ may be obtained by
minimizing the $\chi^2$ difference between the real- and redshift-space
correlation functions. For the fully sampled case this yields $f=1.6\pm 0.2$
and, for the one-in-six redshift space sampling, it yields $f=1.7\pm0.4$.
The quoted errors are the average of the errors  determined assuming maximum
and minumum correlation between data points.

For the QDOT data, the correlation function in real space may be
derived from the
cross-correlation function of the redshift sample with its fully
sampled parent catalogue, computed as a function of projected separation
$\sigma$. Saunders \etal (1992) calculated an integral over $\xi_r$,
$$
\eqalignno{
\Xi(\sigma) &= \int_{-\infty}^\infty \xi_r\left(\sqrt{y^2 + \sigma^2}\right)
\;dy,&(11) \cr
}
$$
in logarithmic bins in $\sigma$.
If $\xi_r$ is approximated as a set of steps,
$\xi_r(r) = \xi_i$ for $s_i<r<s_{i+1}$, we can integrate equation (11)
to give:
$$
\eqalignno{
\Xi_i &= 4/3\Bigl [\xi_i (s_{i+1}^2 - s_i^2)^{3/2} \cr
      &+ \sum_{j=i}^\infty  \xi_j ( (s_{j+1}^2
- s_i^2)^{3/2} - (s_{j}^2 - s_i^2)^{3/2} - (s_{j+1}^2 - s_{i+1}^2)^{3/2} +
(s_{j}^2 - s_{i+1}^2)^{3/2})\Bigr ]\ .&(12) \cr
}
$$

The redshift-space analogue of $\Xi(\sigma)$ [denoted by $\Xi_s(\sigma)]$
can be calculated numerically from equation (11) and our estimates of
$\xi_s$ for the QDOT sample. We assume $\xi_s(s)=0$ for $s>50 {\rm
h}^{-1}$ Mpc and estimate errors by splitting up the survey into octants
and measuring the scatter in the estimate of $\Xi_s$ for each octant.

In Figure~4a we compare the resulting $\Xi_s$ with the Saunders \etal
estimate of $\Xi_r$. The two estimates are in good agreement and show no
evidence for any distortion of the correlation function measured in
redshift space.  To quantify this, we determine $f$  in equation~(10) as
above, by
minimising the $\chi^2$ difference between $\xi_s$ and $f \xi_r$ in the
range $5-30 {\rm h}^{-1} $ Mpc. This gives $ f = 0.8 \pm 0.5 $ ($1 \sigma$
error). A
possible  concern with this approach is that $\Xi_s$ is an integral over
$\xi_s$ to large distances (formally to infinity) and the errors in $\xi_s$
at large separations can introduce correlated errors in $\Xi_s$ on all
scales. We can check whether this is a problem by comparing our
measurements of  $\xi_s$ with estimates of $\xi_r$ derived by
differentiating $\Xi$ numerically. This comparison is carried out in
Figure~4b. As before, we estimate uncertainties from the results for each
octant. Differentiation introduces noise into $\xi_r$, but the final answer
for the ratio of amplitudes is consistent with our previous estimate, {\it
i.e.} $f= 1.0 \pm 0.4 $ ($1 \sigma$ error) over the range
$5-30 {\rm h}^{-1} $ Mpc. (Similar results are obtained if the fits are done
over the range $10-30 {\rm h}^{-1}$ Mpc.)
We adopt this value as our final answer. This result sets a $2 \sigma$ {\it
lower} limit of $b_{IRAS}/\Omega_0^{0.6} > 1.05$, close to the value
$b_{IRAS}/\Omega_0^{0.6}=1.2\pm0.2$ determined by comparing measurements
of the peculiar velocity of the Milky Way, and other galaxies, with the
motions inferred from the spatial distribution of QDOT galaxies
(Rowan-Robinson \etal 1990, Kaiser \etal 1991). A similar limit
on $b/\Omega_0^{0.6}$ has been derived from the lack
of redshift-space distortion in the Stromlo/APM redshift
survey of optically selected galaxies (Loveday \etal 1992a).

The are two reasons why our measurement of $f$ is so uncertain. Firstly, we
are attempting to measure small differences between the much larger measured
quantities, $1+\xi_s$ and $1+\xi_r$, and, as the artificial catalogues
show, this requires highly accurate estimates.
Secondly, the test makes very great demands on the ``fair sample"
hypothesis: one must have not only a fair sample of superclusters, but a
fair sample of {\it orientations} of superclusters as well. Nevertheless,
our results  suggest that future surveys of tens of thousands of galaxies
should lead to constraints on the amplitude of mass fluctuations determined
from the clustering distortion which are at least as useful as present
constraints derived from streaming velocities.

\vskip 20pt
\parindent=0pt
{\bf \S 4. LARGE SCALE INHOMOGENEITIES IN THE DISTRIBUTION OF IRAS GALAXIES}
\vskip 20pt
\parindent=36pt

The counts-in-cells analysis of Efstathiou \etal (1990) showed a large
variance in the counts of QDOT galaxies in cells at a distance $\approx
100 {\rm h}^{-1}$ Mpc. Maps of the galaxy distribution show several large
superclusters at this distance, including the prominent Hercules structure
(Saunders \etal 1991, Moore \etal 1992). To investigate this further, we
have estimated $\xi_s$ from the QDOT survey in a series of concentric shells of
increasing distance from the observer. Figure 5 shows the correlation
function of galaxies measured in three shells of radii $5-50 {\rm h}^{-1}$ Mpc,
$50-100 {\rm h}^{-1}$ Mpc and $100-300 {\rm h}^{-1}$ Mpc, using the two
different estimators in equation~2. There are $561$, $636$
and $723$ galaxies in these shells, with median luminosities $2.1\times
10^9h^{-2}L_\odot$, $8.6\times 10^9h^{-2}L_\odot$, and $3.4\times
10^{10}h^{-2}L_\odot$ respectively.

On scales $\lsim 10 {\rm h}^{-1}$ Mpc there is no apparent difference
between the clustering properties of galaxies as a function of shell
distance. On larger scales there is a systematic increase of clustering
strength with shell distance and it is clear that most of the clustering
signal at separations $\gsim 10 {\rm h}^{-1}$ Mpc comes from the most
distant shell ($s \ge 100 {\rm h}^{-1}$ Mpc). This result is independent of
our choice of estimator for $\xi_s$, although Hamilton's estimator gives
somewhat stronger clustering in the nearest shell at large pair separations.
With the present data set
we are unable to determine whether the apparent increase in clustering
strength with shell distance is due to sampling fluctuations in the
different volumes or whether it is due to a dependence of the shape of the
correlation function on luminosity.

To investigate the possibility of luminosity bias further, we have repeated
the projected correlation analysis of Saunders \etal (1992), this time
looking for luminosity dependence. We split the QDOT survey into four
luminosity bins at
$\log(L/h^{-2}L_{\odot}) <9.25,9.25\--9.75,9.75\--10.25,>10.25$, and find
the projected cross-correlation function with the fully sampled parent
catalogue in each case.  The results are shown in Figure 6. To quantify
any dependence of the amplitude of the correlation function on $L$, we
fitted the data points in the range $5-30 {\rm h}^{-1} $ Mpc with a power-law
of the form, $(s/s_0')^{\gamma'}$, where $s_0'$ depends on the median
luminosity in each luminosity interval:
$$
\eqalignno{
s_0' &\propto  L_{med}^\alpha,  \quad  \gamma = {\rm const}\  .&(13a) \cr}
$$
The resulting best fit is
$$
\eqalignno{
 \alpha &= 0.033\pm0.025 \quad (1 \sigma\;\; {\rm error}). &(13b) \cr}
$$
This statistic provides a constraint on the variation of the amplitude of
$\xi(s)$ with luminosity, giving a $2\sigma$ upper limit of a 20\%
increase in $s_0'$ for each decade increase in luminosity. Although the
cross-correlation function is less sensitive than the autocorrelation
function to variations in clustering strength with luminosity, our results
show that much of the variation of $\xi_s(s)$ with shell radius seen in
Figure 5 must arise from sampling fluctuations rather than luminosity
dependence. Our data are nevertheless consistent with a weak dependence of
the clustering amplitude with luminosity.

\vskip 20pt
\parindent=0pt
{\bf \S 5. THE CLUSTER-GALAXY CROSS-CORRELATION FUNCTION}
\parindent=36pt
 \vskip 20pt

The prominent galaxy clusters apparent in optical maps are far less
conspicuous in maps of IRAS galaxies ({\it e.g.} Figure~1 in Saunders \etal
1991). We can quantify this difference by comparing the cross-correlation
function, $\xi_{cg}(s)$, of Abell clusters with IRAS and optical galaxies
respectively. The bright tail in the luminosity function of IRAS galaxies
gives rise to a population of high redshift galaxies in the QDOT survey
which overlaps samples of Abell clusters. As a result, $\xi_{cg}(s)$ can be
estimated in a fairly straightforward fashion. We use the sample of $822$
Abell clusters with measured redshifts in the compilation of Abell \etal
(1989), which has a median depth of $\sim 300 {\rm h}^{-1}$ Mpc and divide
the clusters into different (optically determined) richness classes. The
subsamples contain 467 clusters of richness $R\ge 1$ and 209 clusters of
richness $R\ge 2$.

We estimate $\xi_{cg}(s)$ using the analogue of equation~(2a),
$$
\eqalignno{
{1 + \xi_{cg}(s)} &= {CG(s) \over CR(s)}\biggl({N_R \over N_D}\biggr) \ ,
&(14)\cr
}
$$
where $CG(s)$ is the weighted pair count of Abell clusters lying in an
interval of separation centred on $s$ from a QDOT galaxy, and $CR(s)$ is
an analogous weighted count of pairs consisting of a cluster and a point
chosen from a large random catalogue (containing $10^5$ points) with the
same sky coverage and mean redshift distribution as the QDOT survey. We
use a similar weighting scheme to that described in Section 2, with number
densities, $N_R$ and $N_D$, given by equation~(5). To calculate $J_3$
we adopt the cluster\--galaxy cross correlation function measured by Lilje
and Efstathiou (1988).

The filled circles in Figure 7 show our estimate of the cross-correlation
function, $\xi_{cg}(s)$, between the full sample of Abell clusters and the
QDOT survey. The $1\sigma$ error bars shown were obtained by estimating
$\xi_{cg}(s)$ for four independent subsamples of the data and are quite
similar to the Poissonian errors of the J3-weighted pair counts.
Beyond $3 {\rm h}^{-1}$ Mpc, the cluster-galaxy cross-correlation functions are
well fit by a power-law of the form $\xi_{cg}(s)=(s/s_\circ)^\gamma$. For
clusters with richness $R\ge 1$ we find $s_\circ=10.10\pm0.45 {\rm h}^{-1}$ Mpc
and $\gamma=-1.75\pm0.10$, consistent with the results of Mo \etal (1993)
determined from our data. The amplitude of the cross-correlation is higher
for the $R\ge 1$ than for $R\ge 0$ clusters, but there is no detectable
increase in amplitude for higher cluster richness. Table~1 summarises these
results using formal $\chi^2$ fits to the data beyond 3\hmpc.

At smaller separations, 1.5\hmpc $\le s \le$ 3\hmpc, the number of clustered
QDOT galaxies drops. The shape of $\xi_{cg}$ on small scales is affected by
a combination of velocity errors and peculiar motions of galaxies in
clusters which smear out intrinsic correlations. (This is analogous to the
smearing of the autocorrelation function described in Section (2)).
Modelling these effects requires knowledge of the velocity distribution
function and can, in principle, be performed using the techniques discussed
in Section 2 in connection with the galaxy - galaxy correlation function.
For example, assuming a Gaussian distribution of cluster galaxy
velocities of width $\sigma=800$\kms\ (close to the median 1-D velocity
dispersion for $R\ge1$ Abell clusters), we find a best fit power-law  with
parameters, $s_\circ=7.95\pm0.10 {\rm h}^{-1}$ Mpc, and,
$\gamma=-1.56\pm0.38$, for $R\ge1$ clusters.

For comparison with optical galaxies we refer to Lilje and Efstathiou's
(1988) estimate of the projected
cross-correlation function between Abell clusters of richness class $R\ge
1$ and the Lick galaxy counts (Shane and Wirtanen 1967). These authors
estimated
$\xi_{cg}(s)$ by inverting a Limber-like equation and found a power-law
behaviour for $\xi_{cg}(s)$ with exponent $-2.21\pm0.04$ and clustering length,
$s_\circ=8.8\pm0.6 {\rm h}^{-1}$ Mpc. This result is plotted as a dashed
line in Figure~7. Beyond $s\simeq 3 {\rm h}^{-1}$ Mpc, the cross-correlation
functions for both optical and IRAS galaxies with clusters of richness $R\ge1$
have comparable amplitudes, but different slopes.
At small separations, $\xi_{cg}$ for optical galaxies
continues to increase at separations $\lsim 0.5 {\rm h}^{-1}$ Mpc, and
rises significantly above $\xi_{cg}$ for QDOT galaxies. Note that
the QDOT counts in this regime are affected by velocity broadening.
Nevertheless, a difference in the relative distributions of
optical and IRAS galaxies near the centres of rich clusters is expected
from Dressler's (1980) morphology-density relation.

\vskip 20pt
\parindent=0pt
{\bf \S 6. DISCUSSION AND CONCLUSIONS}
\vskip 20pt
\parindent=36pt

We have carried out a clustering analysis of the QDOT survey of IRAS
galaxies, focussing on three specific issues: $(i)$ the form of the
two-point correlation function on large scales and its dependence on galaxy
luminosity, $(ii)$ the amplitude of the redshift-space distortion of the
correlation function and $(iii)$ the relative distributions of IRAS and
optical galaxies around Abell clusters.

At pair separations less than 25\hmpc, the
two-point correlation function of QDOT galaxies is well approximated
by a power-law of the form $\xi(s)=(s/s_\circ)^\gamma$, with $\gamma =
-1.11\pm 0.09$ and clustering length, $s_\circ=3.87 \pm 0.32 {\rm h}^{-1}$ Mpc.
(These numbers take into account the smearing due to small scale peculiar
velocities and measurement errors, and are unaffected by luminosity/density
evolution of IRAS galaxies.)
Beyond $s\simeq 40 {\rm h}^{-1}$ Mpc, our
data are consistent with $\xi(s)=0$. This result agrees well with the
correlation function determined for two other redshift surveys of
IRAS galaxies, the 2 Jy survey of Strauss \etal (1990) and the 1.2 Jy
survey of Fisher \etal (1994).  IRAS galaxies appear to be slightly
less strongly clustered than optical galaxies on small scales. The
difference decreases on larger scales reflecting the shallower slope of the
IRAS autocorrelation function. This is consistent with the
results of Loveday \etal (1992a), who find that the variances in the
counts of optical and IRAS galaxies are similar, $\sigma^2_{opt}/
\sigma^2_{IRAS} = 1.2 \pm 0.3$, on scales $\simgt 20$ \hmpc.

We can compare our results with the analyses of Efstathiou \etal (1990) and
Saunders \etal (1991), who give the variance of counts-in-cells in the QDOT
survey as a function of cell-size, using two different estimation
procedures. This variance is related to the correlation function by a double
integral over the cell volume $V$:
$$
\eqalignno{
\sigma^2 & = \int_V \xi(s_{1,2})dV_1dV_2/V^2 \ .
&(15)\cr
}
$$

In Table 2 we list the results of evaluating this integral over cubical
cells of side $\ell$ using a cubic spline fit to (the uncorrected) $\xi_s$.
The numbers in brackets are approximate errors estimated from the integral
using fits to the top and bottom of the error bars for the $J_3$-weighted
estimate of $\xi_s$ plotted in Figure 2.  (This clearly gives a rather
pessimistic estimate since it assumes that the errors in the estimates of
$\xi_s$ in different bins are 100\% correlated.) The third column of Table 2
gives the counts-in-cells results of Efstathiou \etal In cells of length 30
${\rm h}^{-1}$ Mpc and 40 ${\rm h}^{-1}$ Mpc, our estimated variances are
just within the 95\% errors quoted by Efstathiou {\it et al.}, but in the
other cells the agreement is good. For comparison, in column 4 we list the
counts-in-cells for an optically selected catalogue of galaxies by Loveday
\etal (1992a).
In columns 5 and 6 we give the expected variances for the theoretical
models considered in Section~2. For the standard CDM model we  use the
procedure employed for the real data and integrate the correlation function
determined from our artificial catalogues  (cf Figure~2a). For the
$\Gamma=0.2$ model, we simply integrate the (real-space) linear theory
correlation function. On large scales, the standard CDM model gives variances
which fall outside of the $95\%$ error limits of the data. These results
agree with the analysis of the 1.2 Jy survey by Fisher \etal (1994).

We find no detectable difference between estimates of the QDOT
autocorrelation function in real and redshift space over the range where
we have measured these functions. This is not surprising -- our artificial
catalogues show that the redshift-space distortion
would be detectable in a fully-sampled version of the QDOT catalogue (if
$b=1.5$) but in our sparsely sampled survey the signal is swamped by
statistical errors. Neverthelss, we are able to set a 2-sigma lower limit
of $b^{IRAS}/\Omega^{0.6}>1.05$, consistent with independent
estimates based on studies of the peculiar velocity filed (see for
example, Kaiser \etal 1991, Dekel \etal 1992).

Integrating $\xi_s$ over volume leads to an estimate of
$\sigma_8^{IRAS}=0.58\pm 0.14$ for the {\it rms} count of IRAS galaxies in
top hat spheres of radius $8 {\rm h}^{-1}$ Mpc. This is consistent with
the value, $0.69\pm 0.09$, obtained by Saunders \etal (1992) from an
analysis of clustering in real space. Since $\sigma_8^{opt}$ for optical
galaxies is close to unity, taking the weighted mean of the two IRAS
determinations, we infer a relative ``bias factor", $b^{IRAS}/b^{opt}
\equiv \sigma_8^{IRAS}/\sigma_8^{opt} = 0.65$ on scales of $\sim 8$ \hmpc,
with an error of about $25 \%$.

The visual appearance of the galaxy distribution in the QDOT survey
(Saunders \etal 1991, Moore \etal 1992) shows structures of increasing
apparent richness and overdensity the further out we look. Nearby there
are only a few very rich clusters. At $\sim 70 {\rm h}^{-1}$ Mpc, the
``Great Wall'' (Geller and Huchra 1989) which contains several rich
clusters of galaxies is apparent. Further away, at $\sim 100 {\rm
h}^{-1}$ Mpc, there are the most prominent superclusters in the survey,
such as Hercules, Aquarius-Capricorn and Horologium. This visual
impression is reflected in our quantitative estimates of clustering. For
example, the variance of counts-in-cells is largest for cells lying $\sim
100 {\rm h}^{-1}$ Mpc away.  Consistent with this, our estimates of the
correlation function in concentric shells show a marked increase in
clustering strength with shell radius. Most of the ``large-scale power"
seems to come from structures at distances of $\sim 100 {\rm h}^{-1}$ Mpc.

The apparent increase of clustering strength with distance may be
interpreted in at least two ways. It may be due to sampling fluctuations
of a distribution with an underlying universal clustering pattern, or it
may be due to genuine variations of the correlation function with, for
example, galaxy luminosity. Ideally one might distinguish between these
alternatives by comparing estimates of $\xi$ for galaxies of different
luminosity in the {\it same} volume of space. The shape of the luminosity
function, however, precludes an analysis of this sort for samples of
widely differing intrinsic luminosities selected with a limiting flux. A
slightly different approach, and the one we have followed here, is to
compare the cross-correlation functions of QDOT galaxies of different
luminosities with the {\it projected} parent catalogue of IRAS galaxies.
Although the cross-correlation function is less sensitive than the
autocorrelation function to variations in clustering strength with
luminosity, our results show that most of the apparent increase in $\xi$
with distance is likely to be due to sampling fluctuations.  Nevertheless,
a weak dependence of clustering strength with galaxy luminosity is
consistent with our data, which allow an increase of $\sim 20\%$ in
clustering length for each decade increase in luminosity.

Finally, we investigated the environment of IRAS galaxies by measuring the
cross-correlation of the QDOT survey with a sample of rich Abell clusters.
For separations larger than $3 {\rm h}^{-1}$ Mpc, the mean galaxy count
falls off with cluster radius slightly less steeply than $r^{-2}$. The
cross-correlation amplitude is larger for clusters with (optical) richness
$R\ge 1$ than for clusters with richness $R\ge 0$, but we detect no
further increase for richer clusters. On small scales there is a deficit
of IRAS galaxies relative to optical galaxies. Although this may be
related to Dressler's morphology-density relation, our estimates are
uncertain at such small separations because the cross-correlation function
is smeared out by peculiar velocities and velocity errors. Our results
agree well with those of Strauss \etal (1992) who measured the
distribution of IRAS galaxies around a sample of six nearby clusters.

In summary, the autocorrelation function of the QDOT survey
indicates that IRAS galaxies are less strongly clustered than
optically selected galaxies on scales $\lsim 10$ \hmpc. On larger scales,
the cross-correlations of Abell clusters with optical and IRAS
galaxies, and the similarity of the counts-in-cells variances, suggest
that optical and IRAS galaxies have comparable clustering amplitude.
The large redshift surveys currently under way or at an advanced planning
stage will, in the near future, tighten up some of the results which are
marginal even with a survey the size of QDOT. These include the
redshift-space distortion of the correlation function and the dependence of
clustering strength on galaxy luminosity.

\parindent=0pt
\vskip 15pt
{\bf Acknowledgments}
\vskip 10pt
\parindent=36pt

We would like to thank Karl Fisher and Marc Davis for stimulating
discussions. BM acknowledges support from a NATO/SERC fellowship and the
Center for Particle Astrophysics at Berkeley. Parts of this work have
been supported by the UK Science and Engineering Research Council.
CSF acknowledges a Nuffield Foundation Science Research Fellowship and a Sir
Derman Christopherson
Research Fellowship. WJS acknowledges support from an SERC Advanced
Fellowship.

\vfill\eject

\parindent=0pt
\vskip 20pt
{\bf References}
\vskip 7pt
\parskip=4pt

\pp Abell, G. 1958, {\it Ap.J. Suppl.}, {\bf 3}, 211.

\pp Abell, G., Corwin, H.G.Jr. and Olowin, R.P. 1989, {\it Ap.J.Supp.},
{\bf 70}, 1.

\pp Babul, A. and White, S.D.M. 1991, {\it MNRAS}, {\bf 253}, 31P.

\pp Bean, A.J., Efstathiou, G., Ellis, R.S., Peterson, B.A. \& Shanks, T.,
1983, {\it MNRAS}, {\bf 205}, 605.

\pp Bower, R.G., Coles, P., Frenk, C.S. and White, S.D.M. 1993,
{\it MNRAS}, {\bf 405}, 403.

\pp Davis, M., Efstathiou, G., Frenk, C.S. and White, S.D.M. 1985, {\it
Ap.J.}, {\bf 292}, 371.

\pp Davis, M., Meiksin, A., Strauss, M.A., da Costa, L. and Yahil, A.
1989, {\it Ap.J.Let.}, {\bf 333}, L9.

\pp Dekel, A., Bertschinger, E., Yahil, A.,
Strauss, M.A., Davis, M. \& Huchra, J.P.
1993, {\it Ap.J.}, {\bf 412}, 1.

\pp Dressler, A. 1980, {\it Ap.J.}, {\bf 236}, 351.

\pp Efstathiou, G. 1988, in Proc. Third {\it IRAS} Conf., Comets to Cosmology,
ed. A.Lawrence (Berlin: Springer-Verlag), 312.

\pp Efstathiou, G., Kaiser, N., Saunders, W., Lawrence, A., Rowan-Robinson, M.,
Ellis, R.S.E and Frenk, C.S. 1990, {\it MNRAS}, {\bf 247}, 10p.

\pp Efstathiou, G., Sutherland, W.J. and Maddox, S.J., 1990,
{\it Nature}, {\bf 348}, 708.

\pp Efstathiou, G., Bond, J.R., and White, S.D.M.,
1992, {\it MNRAS}, {\bf 258}, 1p.

\pp Fisher, K.B., Davis, M., Strauss, M.A., Yahil, A. and Huchra, J. 1993,
{\it Ap.J.}, {\bf 402}, 42.

\pp Fisher, K.B., Davis, M., Strauss, M.A., Yahil, A. and Huchra, J. 1994,
{\it M.N.R.A.S.}, {\it 266}, 50.

\pp Frenk, C.S., White, S.D.M., Efstathiou, G. and Davis, M. 1990, {\it Ap.J.},
{\bf 391}, 2.

\pp Geller, M.J. and Huchra, J.P. 1989, {\it Science}, {\bf 246}, 897.

\pp  Hamilton, A.J.S. 1993, {\it Ap.J.}, {\bf 406}, L47.

\pp Kaiser, N. 1987, {\it MNRAS}, {\bf 227}, 1.

\pp Kaiser, N., Efstathiou, G., Ellis, R.S., Frenk, C.S., Lawrence, A.,
Rowan-Robinson, M. and Saunders, W. 1991, {\it MNRAS}, {\bf 252},
1.

\pp Lawrence, A. \etal 1993, in preparation.

\pp Lilje, P. and Efstathiou, G. 1988,
{\it MNRAS}, {\bf 231}, 635.

\pp Loveday, J., Efstathiou, G., Peterson, B.A. and Maddox, S.J. 1992a,
{\it Ap. J.}, {\bf 400}, L43.

\pp Loveday, J.,, Peterson, B.A.,
 Efstathiou, G. and Maddox, S.J. 1992b,
{\it Ap. J.}, {\bf 390}, 338.

\pp Lynden-Bell, D., Faber, S.M., Burstein, D., Davies, R.L.,
Dressler, A., Terlevich, R.J. and Wegner, G., 1988, {\it Ap. J.}
{\bf 326}, 19.

\pp Maddox, S., Efstathiou, G., Sutherland, W. and Loveday, J. 1990,
{\it MNRAS}, {\bf 242}, 43P.

\pp Mo, H.J., Peacock, J., and Xing, X. 1993, {\it MNRAS},
{\bf 260}, 121.

\pp Moore, B., Frenk, C.S., Weinberg, D.H., Saunders, W., Lawrence, A.,
Ellis, R.S., Kaiser, N., Efstathiou, G. and Rowan-Robinson, M.,
1992, {\it MNRAS}, {\bf 256}, 477.

\pp Rowan-Robinson, M., Lawrence, A., Saunders, W., Crawford,J., Ellis,
R.S., Frenk, C.S. Parry, I., Xiaoyang, X., Allington-Smith, J., Efstathiou,
G., and Kaiser, N. 1990, {\it MNRAS}, {\bf 247}, 1.

\pp Rowan-Robinson, M., Saunders, W., Lawrence, A. and Leech, K.L.
1991, {\it MNRAS}, {\bf 253}, 485.

\pp Saunders, W., Rowan-Robinson, M., Lawrence, A., Efstathiou, G.,
Kaiser, N., Ellis, R.S. and Frenk, C.S. 1990,
{\it MNRAS}, {\bf 242}, 318.

\pp Saunders, W., Frenk, C.S., Rowan-Robinson, M., Efstathiou, G.,
Lawrence, A., Kaiser, N., Ellis, R.S., Crawford, J., Xiao-Yang, X. and
Parry, I. 1991, {\it Nature}, {\bf 349}, 32.

\pp Saunders, W., Rowan-Robinson, M. and Lawrence, A. 1992,
{\it MNRAS}, {\bf 258}, 134.

\pp Shane, C.D and Wirtanen, C.A. 1967, {\it Publ. Lick Obs.}, {\bf 22}, 1.

\pp Silk, J. 1992, {\it Australian Journal of Physics}, {\bf 45}, 437.

\pp Smoot, G. \etal {\it Ap.J.Lett.}, {\bf 396}, L1.

\pp Strauss, M.A., Davis, M., Yahil, A. and Huchra, P.J. 1990, {\it Ap.J.}
{\bf 361}, 49.

\pp Strauss, M.A., Davis, M., Yahil, A. and Huchra, P.J. 1992, {\it Ap.J.}
{\bf 385}, 421.

\pp Strauss, M., Cen, R. and Ostriker, J.P. 1993, {\it Ap. J.}, {\bf 408}, 389.

\pp Struble, M.F. and Rood, H.J. 1987, {\it Ap.J.Suppl.}, {\bf 63}, 543.

\pp Vittorio, N., Juszkiewicz, R. and Davis, M. 1986,
{\it Nature}, {\bf 323}, 132.

\pp Vogeley, M.S., Park, C., Geller, M.J. and Huchra, J.P. 1992, {\it
Ap.J.Let.},
{\bf 395}, L5.

\pp White, S.D.M., Efstathiou, G. and Frenk, C.S., (1993) {\it MNRAS}, {\bf
262}, 1023.

\pp White, S.D.M., Frenk, C.S., Efstathiou, G. and Davis, M. (1987) {\it Ap.
J}, {\bf 313}, 505.

\vfil\eject

\parindent=0pt
{\bf Figure captions}
\parskip=10pt

{\bf Figure 1} \ (a) The average redshift-space correlation function for 5
artificial QDOT catalogues constructed from CDM N-body as discussed in the
text. The different symbols correspond to different weighting schemes:
constant unit weighting (filled triangles); equation~(3) with $4\pi
J_3=3000$ (open squares); and equation~(3) with $J_3$ calculated by
integrating the linear theory prediction for $\xi(s)$ with $\Gamma=0.5$
(open circles) and $0.2$ (filled circles). The dashed line shows the linear
theory prediction for the CDM ($\Gamma=0.5$) correlation function. (b) The
$J_3$-weighted ($\Gamma=0.5$) estimate of the CDM correlation function with
$1\sigma$ error bars calculated from the scatter between 5 separate
artificial QDOT catalogues. The error bars to the left and right of each
point are the Poisson errors of the $J_3$-weighted pair counts and the
$1\sigma$ error from 100 bootstrap resamplings of one of the
catalogues respectively.

{\bf Figure 2} \ (a) Two-point correlation functions for the QDOT redshift
survey and artificial catalogues. The filled circles give the QDOT estimate
obtained from equations~(2b) and (3), assuming the $\Gamma=0.2$ linear
theory prediction for $J_3(s_{ij})$. Error bars show $1\sigma$ errors
calculated from the $J_3$-weighted pair counts. (These are similar to the
errors  found by dividing the survey into four quadrants.) The open circles
show the result of allowing explicitly for the effect of number density
evolution.  The dot-dashed line gives the correlation function corrected
for small scale peculiar  velocities and velocity measurement errors,
obtained by fitting the dot-dash line to the data. The stars give the
predictions from the standard CDM model obtained from the N-body simulations
illustrated in Figure~1a, while the dashed line gives the (real-space) linear
theory prediction for $\Gamma=0.2$. (b) Two-point correlation functions
estimated as in (a) for three subsamples of the QDOT survey, volume limited
at $40 {\rm h}^{-1}$ Mpc, $80 {\rm  h}^{-1}$ Mpc and $120 {\rm h}^{-1}$ Mpc
respectively. The filled and open circles give estimates for the total
sample, ignoring and including number density evolution, as in (a), but
using the estimator of equation~(2a).

{\bf Figure 3} \ Two-point correlation functions in real and redshift
space for an ensemble of five $b=1.5$ catalogues constructed from N-body
simulations. The filled and open squares show the mean functions in real
and redshift space respectively for fully-sampled catalogues. The open
circles show the mean function in redshift space for the corresponding
1-in-6 catalogues. (For clarity, these have been offset slightly to the
left.) The error bars represent 1$\sigma$ errors derived from the scatter
in the five independent estimates and agree well with the mean of the
Poisson errors derived from the pair counts. The lower curve is a fit to
the real-space data while the upper curve shows the result of shifting this
same curve vertically by the factor 1.53 implied by equation~(10).

{\bf Figure 4} \ (a) The real-space and redshift-space projected correlation
functions, $\Xi_r$ and $\Xi_s$, for QDOT galaxies. $\Xi_r$ is taken from
Saunders \etal 1992. Error bars on $\Xi_s$ are derived from the scatter between
octants. (b) as (a), but for $\xi_r$ and $\xi_s$.

{\bf Figure 5} \ Two-point correlation functions for QDOT galaxies in
three concentric shells of radii 5-50 ${\rm h}^{-1}$ Mpc (triangles),
50-100 ${\rm h}^{-1}$ Mpc (squares) and 100-300 ${\rm h}^{-1}$ Mpc
(circles). The filled symbols are calculated using the Davis and Peebles
(1982) estimator, equation~(2a), and the open symbols are calculated
using Hamilton's (1993) estimator, equation(~2b). The error bars give
$1\sigma$ errors calculated from the $J_3$-weighted pair counts.

{\bf Figure 6} \ The projected cross-correlation function between the QDOT
survey and its fully sampled parent catalogue split by luminosity. The
luminosity bins considered correspond to $\log(L/h^{-2}L_{\odot})
<9.25,9.25-9.75, 9.75-10.25,>10.25$. Error bars are $1\sigma$ from the
$J_3$-weighted pair counts. For clarity, the points have been offset
slightly in the horizontal direction.

{\bf Figure 7} \ The Abell cluster- QDOT galaxy cross-correlation function.
Results are given for clusters of Abell richness $R\ge 0$ (solid circles),
$R>0$ (open squares) and $R>1$ (open triangles). The $1\sigma$ error bars
were derived by splitting the QDOT survey into four quadrants. The dashed
line shows a fit to Lilje and Efstathiou's (1988) estimate of the
cross-correlation function between Abell clusters and optically selected
galaxies.

\vfill\eject
\bye